\documentclass[a4,12pt]{article}
\usepackage[utf8]{inputenc}
\usepackage{titlesec}
\titleformat{\section}{\bf\large\center}{}{0em}{\thesection{.}\hspace{0.6em}}
\titleformat{\subsection}{\bf\large\center}{}{0em}{\thesubsection{.}\hspace{0.6em}}
\usepackage{url}
\usepackage{amsmath}
\usepackage{amsfonts}
\usepackage{amssymb}
\usepackage{txfonts}
\usepackage[T1]{fontenc}
\usepackage[font={small}]{caption}
\usepackage[margin=1.3in]{geometry}
\usepackage{float}
\usepackage{braket}
\usepackage{marginnote}
\usepackage[sort,authoryear,round]{natbib}
\setcitestyle{aysep={,}}

\usepackage{comment}
\usepackage{tikz}
\usepackage{pdfpages}
\usepackage[hidelinks]{hyperref}
\usepackage[hang]{footmisc}

\title{\textbf{Holistic Versus Fragmented Multiverses}:\\ Empirical Access via Causal and Grounding Signatures}
\author{Baptiste Le Bihan\thanks{University of Geneva, Department of Philosophy,
 Rue De-Candolle 2, CH-1205 Genève. E-mail: \texttt{baptiste.lebihan@unige.ch}}}
\date{November 2026, second version for the \\ \textit{Blackwell Companion to the Philosophy and the Multiverse} \\ (K. Kraay and D. Rubio, eds.)}

\begin{document}
\maketitle

\begin{abstract}
\noindent
Several objections have been raised against the scientificity of multiverse scenarios. This chapter defends the scientificity of certain multiverse scenarios by distinguishing fragmented from holistic multiverses. Concepts of multiverse used in physics are almost always holistic: they embed universes within a broader, unifying physical or metaphysical multiversal structure that can, in principle, leave empirical signatures inside the universes. I develop a classification of such signatures and show how it applies to a number of scenarios from quantum theory, cosmology, and string theory. The framework clarifies why objections such as the `this universe' objection and a newly articulated generalization, the epistemic isolation objection, fail against scientifically motivated multiverses.
\end{abstract}

\newpage

\tableofcontents
\bigskip

\newpage
\section{Introduction}

The idea that reality could comprise a plurality of universes has emerged as a significant theme in both theoretical physics and philosophy.\footnote{This chapter is a long-overdue publication of ideas I first presented in a series of talks beginning in 2021. I am grateful to the many audiences whose questions and comments helped me refine the content of this chapter over the last years, and in particular to participants at the \textit{Multiverse and Philosophy Conference} (June 16–18, 2025) organized by Klaas Kraay and Daniel Rubio. I also thank Emily Adlam, Jacob Barandes, Isaac Wilhelm, and Christian Wüthrich for their feedback, and Neil A. Manson for several especially helpful discussions of the `this universe' objection, useful feedback as well as for introducing me to the basics of baseball. Elsevier's LeapSpace AI tool has been used for research, and OpenAI's ChatGPT for editing and discussing possible objections.} In physics, for instance, eternal inflation posits the endless production of bubble universes, while Everettian quantum mechanics interprets the evolution of the wavefunction as a continual branching into multiple quasi-classical worlds (\citealp{guth2007eternal, Saunders2010-SAUMWE-3, wallace2012emergent}; \citealp{Carroll2019}). Multiverse scenarios also figure prominently in discussions of fine-tuning and anthropic reasoning, sometimes in response to theological design arguments (see, e.g., \citealp{hacking1987inverse}; \citealp{white2003fine}; \citealp{isaacs2022multiple}; \citealp{manson2022cosmic}; \citealp{Saad2024-SAAFSM-2}; \citealp{Ruyant2025-RUYIBB}; \citeauthor{MeachamForthcoming-MEAFMU}, this volume). 

The question of whether we should \textit{ultimately believe} that we live in a multiverse---or even whether we could possibly \textit{know} that we do---remains far from settled. Philosophical treatments of multiverse epistemology have often been framed in Bayesian terms, or more generally as a priori analyses of how rational credences should be updated given anthropic evidence or fine-tuning considerations (see, e.g., \citealp{hacking1987inverse}; \citealp{white2003fine}; \citealp{friederich2021multiverse}). My approach can be read as a counterpoint to Bayesian approaches: instead of using a priori coherence and evaluation, I adopt a \textit{naturalistic metaphysical perspective tied to scientific practice}. It investigates the explanatory role of multiverse hypotheses within scientific practice, and their relation to worldviews. From this vantage point, what matters is not a priori probability assignments but the metaphysical profile of a multiverse, and whether that profile allows for empirical signatures.

Beyond the difficulties surrounding the effectiveness of fine-tuning arguments, many critics contend that multiverse hypotheses are unscientific, either untestable or committed to entities beyond possible observation \citep{ellis2011does, ellis2014scientific}. For example, George Ellis insists that:

\begin{quote}
We need some kind of causal contact with whatever entities we propose; otherwise, there are no limits. The link can be a bit indirect. If an entity is \textit{unobservable but absolutely essential for properties of other entities that are indeed verified}, it can be taken as verified. But then the onus of proving it is absolutely essential to the web of explanation. The challenge I pose to multiverse proponents is: Can you prove that unseeable parallel universes are vital to explain the world we do see? \textit{And is the link essential and inescapable?} (\citealp[43]{ellis2011does}, my emphasis)
\end{quote}

Three distinct objections are often run together in both public and specialist discussions of multiverse hypotheses, namely that (i) they posit unobservable entities; (ii) they lack \emph{current} empirical support; and (iii) they could not, even \emph{in principle}, receive empirical support (see, e.g., \citealp[184]{hossenfelder2018lost}).

These are not equivalent. 

The first objection is not compelling as science routinely posits unobservable entities as long as these posits provide distinctive explanatory and predictive success. Any unobservable entity is, in principle, escapable---provided one is willing to accept the explanatory costs and depart from established scientific orthodoxy. For example, in a Duhemian spirit, one could have resisted commitment to atoms by treating Brownian motion, diffusion, and the gas laws as independent phenomenological regularities. However, this sort of strategy is \textit{ad hoc} and explanatorily costly \citep{psillos2011moving}. Asking for absolute inescapability thus sets an unusually (unreasonable?) high bar, rarely imposed in actual scientific practice (see, e.g., \citealp{carroll2019beyond} for a Bayesian approach).

The more serious worry is (iii). Earlier discussions of the possibility of empirically confirming multiverse scenarios can be found in \citet{friederich2021multiverse} and \citet{read2021landscape}. Friederich, after a nuanced overall assessment, adopts a pessimistic stance, concluding that predictions would be `hopelessly unreliable' \citep[179]{friederich2021multiverse}.\footnote{``It may be, in principle, possible to extract concrete nontrivial predictions from specific multiverse theories using, notably, considerations about self-locating belief. But the researchers'
degrees of freedom that arise in the process---e.g., when making choices about observer proxy, cosmic measure, cosmic background conditions, are prone to be affected by confirmation bias---which makes such predictions hopelessly unreliable'' \citep[179]{friederich2021multiverse}.}  By contrast, in line with my previous work with James Read, I will defend the prospect that we might come to know we live in a multiverse, by identifying specific traces of the multiverse within universes. I will argue that holistic scenarios, understood as those involving multiverse-to-universe determination relations, keep the prospect of empirical confirmation genuinely open.

Section~\ref{beyond} explains why (ii), the absence of \emph{current} confirmation, is a weak objection to multiverse scenarios, and, pulling in the opposite direction, why fine-tuning is at best motivational in the search for new physics. This clears the ground for a positive framework and a nuanced evaluation of the prospects for empirical confirmation. Section~\ref{hol} introduces the distinction between holistic and fragmented multiverses, arguing that only the former admit empirical access in principle. Section~\ref{sig} develops a typology of empirical signatures. Section~\ref{sec:scenarios} applies this typology to cases from quantum theory, string theory, and cosmology. Section~\ref{objection} addresses the ‘this universe’ objection and a generalization I articulate, the epistemic isolation objection. Section~\ref{sec:conc} concludes.

\section{Beyond Fine-Tuning and the ‘Not Testable Now’ Objection}\label{beyond}

This section addresses two distinct possible issues, pulling in opposite directions, before moving on to introduce the framework. First, one could argue that fine-tuning considerations are at best suggestive: they do not supply decisive evidence for a multiverse. They merely show how multiverses might leave empirical signatures (see Section~\ref{sig}). Second, the mere lack of \emph{current} tests is not necessarily an issue: historically, theories were often accepted on explanatory grounds before concrete tests became technologically or even conceptually conceivable.

Fine-tuning arguments exploit the fact that sentient life could not have emerged under slightly different parameter values in our physical laws to suggest that these values should be seen as selection effects within a multiverse: any sentient observer must, by necessity, find themselves in one of the rare life-permitting universes. Yet such arguments are difficult to assess. I take them to be suggestive but not decisive.\footnote{See also Friederich (2021, chs. 4-5) for the view that fine-tuning by itself cannot do the evidential work alone; any genuine support must come from independent traces of a multiversal structure.}

My main resistance is that the parameters typically cited as fine-tuned appear in effective theories rather than in fundamental ones. As such, their values could be artifacts of our incomplete description, patching over ignorance of a more fundamental dynamics—one that might be captured by an empirically confirmed theory of quantum gravity. (For accessible overviews aimed at philosophers of why such a theory is needed, see \citet{wuthrich2005quantize} and \citet{crowther2025we}.)

One might object that effective theories are sufficient, at least to a good approximation, for capturing the modal structure of the physical world, including possibilities that involve different parameter values. On this view, the modal structure of physical theories naturally maps onto the structure of the multiverse. After all, models of physical theories are ordinarily taken to track physical possibilities (\citeauthor*{LeBihanForthcoming-LEBPIP}, 2025). In that sense, models with counterfactual parameter values may, \textit{prima facie}, represent physically possible worlds.

However, varying parameter values is arguably not methodologically equivalent to varying boundary conditions. The former amounts to changing the laws of nature, rather than merely changing boundary conditions under fixed laws. Models with varied parameters therefore lie further from the actual world than those differing only in initial or boundary conditions, and whether they represent genuine physical possibilities remains open to doubt.\footnote{For this reason, the effective criterion for representation of physical possibilities proposed by \citeauthor{Baron2025-BARPTA-14} (2026)—that models of physical theories correctly represent physical possibilities, at least in part, within their domain of application, \textit{absent independent evidence to the contrary}---need not extend to parameter variation. And we already do have such independent reasons to be careful about how to interpret the parameters of effective theories: a more fundamental theory of quantum gravity should underlie current effective theories and account both for the existence of their parameters and for the range of their admissible values. This makes me doubtful that fine-tuning arguments grounded solely in effective theories can provide decisive support for multiverse hypotheses.} It may well be that such models represent only epistemic possibilities---scenarios consistent with what we know of the universe---rather than genuine physical possibilities consistent with the laws of nature.

A distinct worry is that multiverse hypotheses are too speculative because they are `not testable now'. But this cannot possibly be a serious worry, since the absence of present tests is compatible with future testability and does not, obviously, undermine the scientificity of these hypotheses.

To see this, consider the heliocentric model. \citeauthor{Copernicus1543} (1543) [1992] already implied stellar parallax as a test in principle, but the effect was undetectable for centuries; indirect support came in 1725 with Bradley’s discovery of stellar aberration, and the first direct measurement of parallax was achieved by Bessel in 1838, nearly 300 years later. In the meantime, heliocentrism gained support from other observations and on explanatory and unifying grounds, long before stellar parallax could be measured \citep{kuhn1957copernican}.

Likewise, inflation was initially motivated by explanatory gains and developed empirical support only subsequently \citep{McCoy2019-MCCEJA}: proposed to address the horizon, flatness, and monopole puzzles, it later predicted the distinctive, almost scale-free pattern of temperature fluctuations in the cosmic microwave background (CMB). Its ability to generate precise predictions about the pattern of fluctuations observed in the CMB is now considered a successful empirical test \citep[7]{friederich2021multiverse}.

Thus, a lack of current testability is not \textit{ipso facto} an issue. It can take decades, sometimes centuries, to connect a theoretical framework to possible observations, or experiments.

Having set aside both an over-reliance on fine-tuning and an overly stringent immediacy requirement for testing, we can now turn to (iii) in a principled form: \textit{Can certain multiverse scenarios be empirically supported—independently of fine-tuning arguments?} More ambitiously, can we ever know whether we inhabit a multiverse? To tackle these questions, I set aside the pre-theoretic, verificationist use of confirmation, on which a claim is confirmed only if it is infallibly established, since it plays little role in empirical theory assessment. I also set aside discussions of so-called non-empirical or meta-empirical confirmation in physics \citep{dawid2013string, Dawid2019-DAWTSO-6, dawid2022meta}, to focus squarely on empirical signatures in a more strictly empirical sense.

\section{Holistic vs Fragmented Multiverses}\label{hol}

To address whether empirical access is possible, we must first clarify what counts as a multiverse scenario, since some versions rule out by definition any direct or indirect access. The term `multiverse' refers indeed to a heterogeneous family of proposals with distinct structures, physical assumptions, and metaphysical commitments. In order to evaluate the possibility of empirical access, a working definition of `multiverse' broad enough to capture the scenarios at stake, but discriminating enough to distinguish cases with different epistemic prospects, is needed. 

A strict definition in terms of necessary and sufficient conditions is in my view difficult, if not impossible, to articulate, given the rich and varied family of existing proposals. Instead, I will adopt a liberal disjunctive criterion. The term `multiverse' applies to any scenario in which at least one of the following holds: (1) reality includes regions governed by laws similar to ours but with different values for important physical parameters; (2) reality includes regions containing counterparts of us sufficiently similar that one might wonder whether they are, in some sense, versions of us.

This disjunctive definition is deliberately liberal: it does not require universes to be absolutely isolated, whether causally or spatiotemporally, and it leaves open the possibility of various kinds and degrees of connection between them. To make sense of this variety of multiverses, I introduce a distinction that will structure the remainder of the chapter. 

\textit{Fragmented multiverses} are those in which universes are maximally isolated, lacking any overarching unifying physical structure in any sense whatsoever. Multiverses of this kind are often invoked in metaphysics, as with Lewis' modal realism \citep{lewis1986plurality}, or more recently in theological contexts \citep{kraay2010theism, Rubio2020-RUBIDO-3}. \textit{Holistic multiverses}, by contrast, embed all universes within a single unified structure—such as a universal wavefunction, inflating field, cyclic dynamics, or spatiotemporal continuity—that can connect them causally, spatiotemporally, or via scientifically tractable non-causal relations (discussed in the next section). In holistic cases, features of our universe are determined by the larger structure, opening the door to possible empirical signatures. By contrast, fragmented cases render their constituent universes epistemically isolated. Sylvia Wenmackers discusses a similar distinction between `bare multiverse theories' and multiverse hypotheses from physics in Chapter N of this volume. The alignment between the two distinctions is important and captures the fact that usually the multiverses found in physics are holistic. However, as I explain below, string theory might provide a possible case of a fragmented multiverse within physics itself, showing that the two distinctions are not necessarily fully equivalent.

Holistic multiverses have two important features. First, certain properties of the whole (the multiverse) exist over and above the properties of the parts (the universes). Second, they satisfy an epistemic transparency thesis: at least some properties of the multiverse must be accessible from within the universes.\footnote{The holistic multiverse does not necessarily presuppose a privileged discrete decomposition of the whole into distinct universes. The multiversal structure could instead be a continuous, non-fragmented entity, with any conceivable separation into universes being necessarily representational. For a presentation and discussion of this ontology of a multiverse without universes in the context of Everettian quantum mechanics, see \citet{glick2024metaphysical}.} 

To make the distinction between holistic and fragmented multiverses concrete, consider two contrasting scenarios. In eternal inflation, universes emerge from a pre-existing inflaton field governed by definite dynamics. The multiverse is here conceived as a generative structure: as the field evolves, it produces individual bubble universes, each with its own parameter values. These universes are not independent, but causally generated as part of the evolution of a single overarching process. The parameters we observe are not arbitrary but reflect the broader structure of the inflating field and the generative process through which universes are formed. That explanatory dependence is what licenses treating this multiverse scenario as holistic.

By contrast, imagine a theistic creation scenario in which God deliberates over whether to create one universe or many and chooses the latter. Such scenarios have been discussed in relation to whether a multiverse might constitute the best of all possible worlds selected by a divine creator (see, e.g., \citealp{kraay2010theism, Monton2010-MONAMT, Rubio2020-RUBIDO-3, Li2023-LITMTT}; \citeauthor{DonahueForthcoming-DONITT-3}, 2025). In this case the multiverse is just the distributional class of created universes, with no broader physical structure unifying them. Their only unifying connection lies in the divine mind. The theological multiverse is, \textit{by definition}, fragmented. More precisely, in such a theistic scenario, positing an additional physical multiverse structure beyond the distributional class of created universes would be logically possible, but redundant. The metaphysical unity of the collection is already grounded in the divine act of creation in such a way that it is unclear what the explanatory benefit would be of introducing a further physical mechanism of metaphysical unification. The theological multiverse might still be construed as holistic if identified with a scientific multiverse; such an identification is \textit{prima facie} implausible, however, since the generative principles diverge sharply: volitional or moral choices in the theological case versus physical laws or mechanisms in the scientific case.

As we will see, the possibility of empirical signatures in any given multiverse scenario depends crucially on which of these conceptions, fragmented or holistic, is in play. 

Now, a common sceptical worry against the very coherence of the concept of multiverse holds that if other universes are causally or spatiotemporally disconnected from ours, then no empirical access, no scientific knowledge  is possible. But this objection applies only to fragmented multiverses. Holistic multiverses, by contrast, are not \textit{absolutely} disconnected: their unifying structures can provide empirical access, either via direct causal access or via indirect non-causal access.

Another natural objection to \textit{holistic} multiverses this time, is that if we had direct observational access to another universe, this so-called `other universe' would simply be absorbed into the definition of our own universe (see, e.g., \citealp{smolin2009unique}). Two clarifications. First, holistic multiverses need not involve direct causal connections at all, and thus no direct observational access; unity can be indirect via a mix of causal and grounding relations, as will be shown in the next section. Second, even when there is causal access, classification is partly semantic and historically contingent. A fruitful comparison is the early-twentieth-century debate over the nature of \emph{island universes}. Until the 1920s, it was unclear whether spiral nebulae were part of the Milky Way or very distant systems of entities, then construed as other universes but later recognized as galaxies beyond our own.\footnote{Henrietta Leavitt’s work on Cepheid variables provided a new empirical method---standard candles---for measuring cosmic distances (see \citealp{Breuval:2025rmc}). Edwin Hubble applied this method in 1923–24, showing that many nebulae lie outside our galaxy \citep{hubble1929relation}.} What were once construed as other universes were reclassified as proper parts of a broader universe.

This analogy is not perfect: galaxies are causally connected to us, whereas universes in a holistic multiverse may not be. But it illustrates that what we describe as `our universe' is partly context-dependent, shaped by the limits of what we can currently observe, and the particular theorisation we make of those limits. If a holistic multiverse of the causal kind were to gain empirical support, we might one day redescribe it as a more expansive conception of \emph{the} universe. That possible semantic shift does not, however, make the current idea of a multiverse otiose: prior to such reconceptualisation, the presently unobservable regions would still instantiate the salient features we naturally associate, presently, with distinct universes.

\section{Empirical Signatures}\label{sig}

This chapter adopts a pragmatic approach to keep the issue tractable: it asks whether multiverse scenarios could receive the same sort of empirical support as other theoretical claims in physics, while bracketing broader debates about the general nature of empirical support and confirmation. To provide minimal structure, I draw on a recent formulation by van Fraassen of what he calls \textit{empirical grounding}. Although van Fraassen is famously skeptical of metaphysics, his account highlights a crucial point: a theory is empirically grounded if the quantities it posits can, at least in principle, be connected to observable outcomes through theoretical links:

\begin{quote}
The requirement of empirical grounding is that, for any quantity occurring in the theory, it must be possible in principle to determine its value:  
a. from direct measurement results,  
b. based on the theoretically posited connections,  
c. and consistent with those connections, subject to:  
d. concordance: every such determination has to yield the same result,  
e. non-triviality: but could have had different results that would have contradicted the theory. \citep[37]{vanFraassen2025}
\end{quote}

\noindent This captures what I mean by an \textit{empirical signature}: not necessarily a direct measurement, but a quantity that can be related to an empirically accessible observation via a chain of theoretical links. Signatures are usually indirect: they rely on theoretical connections that mediate between theory and observation.

Building on this idea, and now moving into metaphysical territory beyond van Fraassen, I distinguish two main kinds of empirically relevant determination relations in the world, which underlie such theoretical links, with further subtypes. I use `determination’ as a general label for dependence relations. These divide into two kinds: causal (nomic) and grounding (non-nomic). Following A.~Wilson’s (\citeyear{wilson2021explanations}) mediation criterion, the contrast turns on a \textit{nomic criterion}: \textit{causation} is determination mediated by dynamical laws of nature, while \textit{grounding} is a non-nomic, metaphysical in-virtue-of relation that is not law-mediated.

\subsection{Direct Causal Signatures}\label{directcaus}

A physical structure leaves a direct causal signature when it interacts directly and causally with observable systems, producing directly detectable traces via causal interaction. The existence of the causal relation in the world allows causal access by epistemic agents. `Direct' here is comparative rather than absolute: Quine–Duhem holism plausibly rules out fully isolated confirmation. The point is that the observed effect is produced by another universe, or the multiverse structure itself, in a way that requires relatively little mediation in the theoretical representation, compared to the grounding cases.

Direct causal signatures admit two subtypes:

\begin{enumerate}
\item \textit{Spatiotemporal causation.} That is the standard case. Causal influence propagates through a spatiotemporal structure. Paradigm cases include signals or collisions of other universes or multiverse structures that imprint observable traces in our past light cone (this could happen in certain models of eternal inflation, see below).

\item \textit{Non-spatiotemporal causation.} Causation need not necessarily be tied to spatiotemporal propagation as soon as we relax the view that causation requires spacetime \citep{Baron2015-BARCST-3}. Note that, logically speaking, certain relata can be spatiotemporal (by being located in spacetime) while the causal relation connecting them is not a propagation in spacetime: consider, for example, Newtonian gravity’s action at a distance. Or, in the context of the possible emergence of spacetime in quantum gravity, one can endorse a causal theory of spacetime according to which the fundamental relations are non-spatiotemporal yet causal. Such fundamental causation could allow apparently isolated universes to be causally connected nevertheless. This could be the case in certain scenarios involving spacetime emergence in quantum gravity as mentioned, but also in Everettian quantum mechanics (see Subsection \ref{emergence}).
\end{enumerate}

\subsection{Indirect Signatures by Grounding}

A theoretical structure leaves an indirect signature by grounding when it \textit{grounds} observable features of our universe. What is observed are the grounded consequences, not the grounding structure itself. Grounding is non-causal and non-nomic, but it can nevertheless be scientifically tractable when it exhibits counterfactual sensitivity: had the grounding structure been different, the observable features would have differed as well \citep{kortabarria2024scientific}. The literature on grounding (e.g., \citealp{Fine2001-FINTQO, Schaffer2009-SCHOWG, rosen2010metaphysical}) provides the template; here it is used in a liberal sense that does not presuppose a single primitive `Big-G' relation, but could range over a number of distinct relations (cf.\ J.\ \citealp{Wilson02112014}).

Again, one can distinguish between two sorts of signatures by grounding:

\begin{enumerate}
\item \textit{By local grounding.} Features of our universe are explained by relatively local structural facts within a larger multiversal base. In Everettian quantum mechanics, for instance, decoherence does not entirely eliminate interference between branches. Residual interference, although extremely small, could in principle leave an empirical signature. Importantly, interference should not necessarily be understood as a genuine, causal interaction. Such effects could rather be analyzed as locally grounded in the relevant structure of the more fundamental universal quantum state, not as causal relations between branches. This is a natural consequence of the view that the multiverse is emergent \citep{wallace2012emergent}. Note that in what follows, the expression `local' should not be interpreted as necessarily meaning `spatial' or `spatiotemporal' location; location can be construed in a more liberal sense allowing non-spatiotemporal location in possibly non-spatiotemporal structures \citep[100–101]{le2018priority}.

\item \textit{By global grounding.} Certain features of our universe can be explained in virtue of statistical properties of the multiverse as a whole. In Everettian quantum mechanics, for example, global properties of the universal quantum state could determine which statistical patterns are typical across branches (consider, for instance, Born-rule frequencies). Such patterns could therefore provide indirect empirical signatures via global grounding. The analogy with general relativity is instructive: global properties of spacetime  are not reducible to purely local features, yet they constrain certain local matters of fact \citep{Lam2007-LAMTSN}. For example, the location of black hole event horizons depends on the structure of the entire spacetime, and the properties of any given `present' slice are fixed in relation to the rest of the spacetime (\citealp{ashtekar2005some}; \citealp{curiel2019many}; \citealp{baron2023trouble}).\footnote{Physicists may object that horizons are defined in terms of global causal structure, hence still causal. More technically, the point here is that the causal structure can serve as the relatum of a determination relation: the grounding of local features in global ones. Note that the analogy is only partial, as the exact location of the event horizon is not directly observable.}
\end{enumerate}

The model of empirical signature by grounding, local or global, is not a new inferential method, but a reinterpretation of familiar scientific reasoning. Consider, for example, Perrin’s work on Brownian motion: the observed motion of suspended particles provided indirect empirical support for the existence of unobservable molecules. Although the collisions between fluid molecules and suspended particles are causal, the macroscopic motion of the latter can be understood as grounded in the collective motion of their microscopic constituents. Grounding earns its epistemic keep here because it is connected to causation in a broader web of dependence relations. What can be observed and confirmed is not only what causes what, but also what grounds what, provided that the grounding relation explains observable features and meshes with an empirically successful theoretical framework.

Now, it might appear that signature by grounding suffers from underdetermination of the theoretical claim by the empirical evidence. If $Y$ is grounded in $X$, the same $Y$ might also be grounded in some alternative $Z$ or $W$. But this is the generic problem of underdetermination, not a special defect of grounding-based signatures. If a candidate ground $X$ systematically explains accessible phenomena $Y$ better than alternatives, and integrates more successfully with surrounding theory, then it can receive empirical support through inference to the best explanation. It would be excessive to require ruling out every possible alternative.

Overall, the taxonomy distinguishes four primary forms of determination that can yield empirical signatures: direct causal signatures via \textit{spatiotemporal} causation; direct causal signatures via \textit{non-spatiotemporal} causation; indirect signatures by \textit{local} grounding; indirect signatures by \textit{global} grounding.

The next section maps this taxonomy onto leading multiverse scenarios. 

\section{Scenarios}\label{sec:scenarios}

This section applies the framework from Section~\ref{sig} to four case studies and reviews: (i) what unifying structure (if any) makes the scenario holistic rather than fragmented; (ii) the available empirical signatures and (iii) what the main empirical/conceptual obstacles are. The aim is not to evaluate their viability, but to show how such approaches connect or could connect multiverse structures to empirically accessible features of our universe. 

\subsection{Everettian Quantum Mechanics}

Everettian quantum mechanics (EQM) takes the Schrödinger equation ontologically seriously, treating it as providing a complete description of the evolution of physical systems. In decoherence-based EQM, the universal wave function evolves unitarily, without ever undergoing collapse. However, interactions between degrees of freedom can simultaneously increase entanglement while suppressing interference between different configurations of those degrees of freedom. As a result, relatively autonomous quasi-classical structures emerge, constituting the branches of the quantum multiverse \citep{wallace2012emergent}. 

The multiverse, identified with the collection of branches, thus emerges from a relatively more fundamental universal quantum state. On this account, the branches are ontologically dependent on the universal state, and thus EQM exemplifies a holistic multiverse: the unity of the whole quantum state determines the properties of the branches/universes.

Although Wallace treats the emergence relation as one of patterns, in Dennett’s sense, another metaphysics-friendly option is to use \emph{grounding} instead. Following A.~\citet{wilson2020nature}, one can take the relation between the universal state and its branches to be a non-causal, asymmetric metaphysical dependence. On this view, empirical data available \emph{within} a branch can justify belief in the existence of features of the broader multiversal structure, not through causal contact with other branches, but because the universal state grounds that branch and its observable properties.\footnote{An issue here is that Wallace's version of EQM construes the universal quantum state as fundamental, with both the multiverse and its branches being emergent. In this context, I will tweak the standard terminology and use `multiverse' as shorthand for the fundamental universal quantum state, rather than the emergent collection of universes.}

In this picture, the multiversal structure, identified with the universal quantum state, could produce empirical signatures via \textit{local grounding}. Decoherence provides an imperfect, approximate autonomy: it does not eliminate cross-branch interference entirely. Residual cross-branch interference, while normally extremely small, could in principle lead to empirical traces. On the Wallace/Wilson EQM picture, however, such phenomena would not amount to literal causal relations between branches. Since branches are emergent structures grounded in the universal quantum state, their residual interference is better understood (although this is not a logical necessity) as a mere manifestation of that more fundamental structure, and not as genuine causal interactions between branches.

Another bumpier road might be provided by \emph{global} properties of the universal state. If these global properties could determine which statistical patterns are typical across branches (consider for instance Born-rule frequencies), then they could perhaps provide an indirect signature of the multiversal structure by global grounding. 

Finally, note that because EQM presupposes exact unitarity, any empirically confirmed departure from Schrödinger dynamics would falsify the EQM multiverse picture outright. Spontaneous-collapse theories could naturally entail such departures by adding stochastic, non-linear terms to the Schrödinger equation, entailing empirical signatures: for instance a universal background of spontaneous X-ray or $\gamma$-ray emissions. The situation looks increasingly bleak for spontaneous collapse models.\footnote{Certain experiments (e.g., the 2025 XENONnT analysis) have already excluded much of the parameter space originally deemed natural for spontaneous-collapse theories \citep{aprile2025csl}.} But this highlights an important point: EQM is directly falsifiable.

The epistemic situation is asymmetric. On the one hand, detection of the predicted collapse signals would refute all strictly unitary interpretations in one stroke, Everettian or otherwise. On the other hand, continued null results cannot single out EQM, but instead further strengthen the family of unitary interpretations. Within this family, EQM affords empirical signatures primarily indirectly and non-causally. Because these indirect signatures are shared with other strictly unitary contenders, they do not single out EQM on empirical grounds alone. 

Whether underdetermination among the unitary options could be broken in the future is an open question, but there is at least one principled route: if EQM were the only unitary approach able to extend coherently to relativistic QFT and, eventually, to quantum gravity. For illustration, \citet{wallace2023sky} argues that no empirically successful \emph{interacting} QFT generalizations currently exist for spontaneous collapse theories, and Bohmian mechanics. Constraints of this kind, whether grounded in QFT or in a theory of quantum gravity, could potentially provide an empirical basis for preferring EQM over its unitary rivals.

In sum, EQM exemplifies a holistic multiverse scenario: the universal quantum state is the unifying structure grounding all branches. Empirical access, in the strict sense, is possible indirectly via local and global grounding signatures. In addition, EQM is open to falsification, since any confirmed departure from unitarity would undermine the entire multiverse picture. This asymmetry highlights both the resilience and the vulnerability of EQM within the unitary family of interpretations.

\subsection{Eternal Inflation}

In contemporary cosmology, the inflationary paradigm was originally introduced to explain several otherwise puzzling features of the early universe: its large-scale smoothness, its uniformity in all directions, and its near-perfect spatial flatness. The idea is that, in its earliest moments, spacetime underwent a phase of extremely rapid expansion. Inflation itself, in contrast to eternal inflation, already faces a number of challenges about testability, model-building, and permanent underdetermination, that I will set aside here (see, e.g., \citealp{dawid2023testability}; \citealp{wolf2025navigatingpermanentunderdeterminationdark}).

What is remarkable is that in wide classes of inflationary models, once inflation begins it does not end, at least not everywhere. Certain regions continue to expand indefinitely, so that, overall, the process never terminates. New bubble universes constantly form within the inflating background, each undergoing its own local big bang and evolving in causal isolation from the rest of the multiverse.\footnote{There are at least two distinct physical routes from inflation to eternal inflation. First, the inflating state could eventually decay, but so slowly that expansion will always outrun the decay, in such a way that the total inflating volume continues to grow. Second, quantum fluctuations could occasionally push the inflaton field back up its potential, thereby nurturing inflation sufficiently often for the overall inflationary process to continue indefinitely.} This is the scenario known as \emph{eternal inflation}. The eternity in play refers to endless continuation into the future; call this property \emph{sempiternity}. Eternal inflation thus posits a sempiternal inflationary process: it has no natural termination in the forward temporal directions, without implying anything about its temporal past. While many working in the inflationary tradition regard eternal inflation as the natural outcome once quantum fluctuations are included (see \citealp{guth2007eternal}), others, most prominently \citet*{ijjas2013inflationary}, dispute this, maintaining that the models consistent with current data do not naturally lead to sempiternity (see \citealp{dawid2023testability}).

Now, suppose inflation is empirically established as the correct description of the early universe. Many already take this to be the case.\footnote{On a broad reading, even Penrose’s Conformal Cyclic Cosmology, discussed below, could be interpreted as an inflationary theory, one in which the inflationary phase occurs before rather than after the big bang.} Second, suppose it is shown that the specific kind of inflation realized in our universe \textit{necessarily} entails sempiternity.\footnote{Here `necessity' should be read in a broad sense: not logical necessity, but the claim that once the relevant inflationary dynamics are in place, eternal continuation follows with overwhelming probability.} Then the same data that confirm inflation locally could, by the structure of the theory, also provide indirect empirical support for eternal inflation. This would be a theoretical prediction rather than an empirical signature (by causation or grounding) since the observations would support a dynamical theory naturally entailing the existence of other universes.

Some models of eternal inflation also make room for direct causal traces. For instance, if another bubble collided with ours in the very early stages of its formation, the impact could leave a distinctive imprint on the CMB or, in some models, a gravitational-wave signal. As \citet[6]{carroll2019beyond} emphasizes, the fact that any subclass of multiverse models can be empirically constrained is already a significant epistemic achievement. Were such a trace ever detected, it would be construed as a strong confirmation. If we are fortunate enough to inhabit a region of the multiverse that carries such a trace, we could place the theory on solid empirical ground. But if we are not lucky in that way, then the absence of a trace need not count against eternal inflation. The situation is analogous to the 1919 solar eclipse: general relativity was spectacularly confirmed by Eddington’s expedition, but had there been no eclipse at the right time and place, that absence would not have counted, obviously, as evidence against Einstein’s theory.\footnote{My thanks to Jacob A. Barandes for raising this point at the Multiverse conference in Toronto.}

Nevertheless, eternal inflation faces a famous difficulty in formulating probabilistic predictions about what observers in an eternally inflating multiverse should expect: the \textit{measure problem}. The measure problem is grounded in the fact that eternal inflation generically produces \textit{infinitely} many universes. To make predictions, say about the probability of a particular value for a fundamental constant inside one universe, one needs a way of stating what a typical bubble is, so that we can ask what observers like us should expect to see, if we are indeed in a typical universe. But with infinitely many universes, the ratios of the sizes of different infinite sets of universes are not uniquely defined according to standard mathematical approaches. Different ways of weighting those infinite sets can make almost any outcome look typical.  

Yet following Carroll (2019), I take the problem to be a technical challenge rather than a necessarily decisive objection. One optimistic scenario, which I find hard to rule out completely, is that the measure need not be stipulated at all, but could instead fall out of the physics. On this view, a future, more fundamental theory of quantum gravity could determine a space of dynamical solutions with relative weights determined by the dynamics themselves.

This is quite a natural idea once we distinguish physical possibility from mere logical possibility: the space of what the laws allow is not coextensive with what we can conceive of, \textit{a priori}. As our understanding of physics deepens, the space of physical possibilities typically shifts, and may well shrink. Since typicality depends on the space of dynamical possibilities and their relative weights, rather than on all \textit{a priori possibilities}, a new theory with a new space of dynamical models could modify what counts as typical.

Eternal inflation thus provides two possibilities for providing empirical support for a multiverse scenario: via (i) direct causal traces, which would deliver confirmation if they occurred in our past light cone, and (ii) indirect support through the theoretical prediction of sempiternity from empirically established inflationary dynamics. Clearly, eternal inflation models are candidates for holistic multiverses, since the inflaton field plays the role of an overarching multiversal structure.

\subsection{String Theory}

String theory is a candidate theory of quantum gravity. It describes a vast multiplicity of consistent dynamical possibilities. In the semantic sense of scientific theories, these are models of the theory: mathematical structures that satisfy its kinematical and dynamical constraints. Each model represents, via its geometry and associated structures, a phenomenological universe by determining its particles, its constants and its effective laws. The sheer abundance of such models, often estimated at numbers as high as $10^{500}$, is what has come to be called the \textit{string landscape}.

The landscape is the mathematical space of vacuum solutions to string theory, following Susskind’s influential definition \citep[574]{Susskind2005}. Note that this is a standard feature of physical theories in general: general relativity, for instance, also admits a vast space of dynamical solutions. The mere existence of many solutions is not yet a threat to empirical predictivity, since, as James Read and I have argued elsewhere \citep{read2021landscape}, we do not even have one string solution demonstrably compatible with all empirical data. There is therefore no established problem of underdetermination among multiple solutions compatible with empirical data.

However, many physicists have been tempted to reify the solution space into a multiverse of concrete universes, corresponding to these mathematical solutions for a number of reasons. Whether the resulting multiverse is fragmented or holistic then depends on the motivations and on the presence (or absence) of a unifying, empirically supported theoretical device that glues the universes together. If the reification is driven by \emph{a priori} considerations---whether theological or scientific in the broad sense of appealing to scientific yet non-empirical principles---it lacks an empirical ground. Such motivations are what Read and I have called `rationalist' \citep{read2021landscape}: for example, the desire to explain away apparent arbitrariness in parameters or initial conditions by insisting that all possibilities must be realized, or more generally by insisting on the principle of sufficient reason, namely that everything must have a reason \citep{unger2015singular}. If the justification is purely grounded in rationalist principles, then the solutions are simply posited side by side, with no overarching structure to unify them.\footnote{There is, of course, no \emph{physical} or otherwise empirically tractable structure. A rationalist might still insist that reality has a purely logical, theological or rational order, something on which epistemic agents can plug directly, without the mediation of the natural world. But such a structure would not qualify as the sort of metaphysical unifying structure that makes a multiverse holistic, in my terminology.}

By contrast, if the reification is tied to \emph{empirical} considerations, the landscape can become physically reified via an empirically motivated mechanism that causally generates or metaphysically grounds the universes. The resulting multiverse is holistic, since the universes are no longer the mere result of reifying the dynamical space of the theory, but generated and explained by an empirically available mechanism. Several candidates for such mechanisms have been proposed:

\begin{enumerate}
\item Eternal inflation can populate different regions of the landscape with bubble universes formed by tunnelling between different vacua, i.e., universes with different geometries and entities (see, e.g., \citeauthor{Susskind2005}, 2005; \citeauthor{tegmark2007multiverse}, 2007; \citeauthor{Freivogel:2011eg}, 2011).
\item Brane cosmology posits a higher-dimensional 10- or 11-dimensional spacetime in which our universe is a 3-brane \citep{papantonopoulos2002brane}. Cosmological phenomena such as brane collisions can generate new universes, and some scenarios even link brane interactions to observable signatures such as dark matter or dark energy.
\item Everettian quantum mechanics can be seen as generating a branching structure in which different branches instantiate different effective vacua, provided one accepts a unification of the quantum and string-theoretic pictures (see, e.g., \citeauthor{PhysRevD.84.105002}, 2011).
\end{enumerate}

Only when coupled to an empirical mechanism does the string landscape multiverse become a genuine candidate for empirical signatures. Absent such embedding of the universes into a broader, empirically accessible structure, the string theory multiverse would remain fragmented.

\subsection{Cyclic Models: Conformal Cyclic Cosmology}

While the notion dates back at least to the Stoic cosmology and the doctrine of ekpyrosis---the destruction and rebirth of the cosmos---modern theoretical physics has revived cyclic models within general relativity and quantum cosmology. Cyclic models regard our observable universe as just one epoch in a succession of temporally ordered universes. Unlike in more standard conceptions of multiverses, the other universes stand in temporal continuity with our own, with the later universes being causally related to earlier ones. This opens the door to direct causal signatures if traces of previous universes survive into the next.

Cyclic models come in a wide variety of kinds (see, e.g., loop quantum cosmology bounce models: \citealp{bojowald2008loop} and \citealp{huggett2018temporal} for philosophical discussion; cyclic brane cosmologies: \citealp{steinhardt2002cosmic}; or Penrose's conformal cyclic cosmology: \citealp{penrose2010cycles}). Viewing cyclic cosmologies as instances of a multiverse scenario relies on an auxiliary assumption, namely eternalism: all of the successive universes co-exist \textit{simpliciter}, and can be existentially quantified over. The succession can happen in time, or in a broader ordering structure similar enough to time to justify the eternalist talk.\footnote{As I have argued elsewhere, eternalism is only contingently tied to time; what matters is that all the proper parts of the considered ontic structure co-exist \textit{simpliciter} \citep{le2020string}.}

I will focus on a particular realization of this idea, namely Penrose’s conformal cyclic cosmology (CCC), in which the remote future of one universe is conformally rescaled to become the Big Bang of the next (\citealp{penrose2006before, penrose2010cycles, penrose2014gravitization}; \citealp{meissner2025physicsconformalcycliccosmology})---see \citet{le2024great} for a presentation aimed at philosophers. This continuity is based on the observation that massless particles, such as photons and gravitons, are insensitive to scale and thus could traverse from one universe to the next without obstruction. CCC predicts that specific observational imprints from the previous universe should be empirically accessible via the CMB. 

Two classes of empirical signatures can be emphasized. First, concentric low-variance circles in the CMB temperature map could arise from violent events such as black hole mergers in the previous universe, whose radiation, associated with massless particles, propagates through the conformal boundary. Second, Penrose predicts the existence of Hawking points: localized bright spots in the CMB resulting from the final bursts of Hawking radiation emitted by supermassive black holes just before their complete evaporation. These predictions are being investigated in current observational cosmology, and the evaluation of the data remains highly controversial \citep{jow2020re, an2020apparent, lopez2021searching, bodnia2024questcmbsignaturesconformal}.

Thus, CCC offers another compelling counterexample to the claim that multiverse hypotheses are inherently untestable. Temporally structured multiverses can, in principle, leave observable traces in our current universe, via causal connections to previous universes. Moreover, CCC aims to provide an explanation for the low-entropy condition of the early universe---the so-called past hypothesis---by arguing that black holes in the late stages of one universe act as entropy sinks, resetting the conditions for the next.

In sum, Penrose’s CCC exemplifies yet another species of holistic multiverse. The succession of universes is governed by a unifying conformal structure that relates the parts, rather than leaving each universe metaphysically isolated. What makes CCC and cyclic models especially distinctive among the class of multiverse scenarios is the temporal continuity between universes: because successive universes are connected along past-directed null paths, observable signals from earlier universes could in principle propagate into ours. While these predictions remain obviously controversial, they illustrate how cyclic cosmologies could in principle admit empirical signatures via causation.

\subsection{Spacetime Emergence}\label{emergence}

Finally, let us describe in more detail the option introduced in Subsection \ref{directcaus}, that non-spatiotemporal causation could directly connect spatiotemporally disconnected universes. In prospective quantum-gravity theories, such as \textit{string theory}, \textit{loop quantum gravity}, and \textit{causal set theory}, spacetime, or at least some of its core properties, are construed as emerging from a relatively more fundamental non-spatiotemporal structure  (for philosophical introductions, see e.g., \citealp{PBS, HUWU}). This suggests an intriguing possibility: spatiotemporal worlds might be \emph{grounded} in a non-spatiotemporal multiversal structure, so that even if universes are spatiotemporally disconnected at the emergent spacetime level, they could still be \emph{causally connected} at the fundamental level by relations within a connected non-spatiotemporal ground.

A concrete illustration comes from \textit{causal set theory}, where the fundamental structure is a discrete partial order over elements that come into existence one by one, without spatial or temporal metrics (for a philosophical introduction, see \citealp{wuthrich2024philosophy, HUWU}). Spacetime distances then emerge from this underlying order. On many readings, causal set theory suggests that the physical world is fundamentally causal \citep{baron2025causal}, exemplifying the more general possibility that spacetime is composed from relatively more fundamental causal relations \citep{baron2022composing, baron2023causal}. If such (admittedly speculative) models are correct, then universes that are observationally and causally disconnected \emph{in spacetime} might nonetheless exert detectable influence \emph{from within} via their joint presence in, and causal relations across, the more fundamental causal structure.

How exactly this might occur remains speculative, but an illustrative example is provided by \textit{brane cosmology} \citep{papantonopoulos2002brane}. In brane cosmology, phenomena such as dark matter or dark energy could potentially be understood as indirect effects of interactions between our observable three-dimensional space and other branes---that is, other universes---situated within a higher-dimensional spacetime. For instance, dark matter, which refers to the source of gravitational effects observed in the absence of expected ordinary observable matter, could result from gravitational waves propagating in a higher-dimensional space, between our three-dimensional universe and other three-dimensional universes. While some interactions, such as electromagnetism, would be confined to individual branes, gravity could extend into the higher-dimensional spacetime, thereby explaining why we observe gravitational effects without corresponding observable matter (in particular, because photons would remain trapped within each brane). Although each three-dimensional brane would be largely shielded from direct interaction with the others because of spatial separation, they could nonetheless remain causally connected via the higher-dimensional spacetime. 

However, the analogy between brane cosmology and spacetime emergence is limited since the higher-dimensional structure of brane cosmology hosting the three-dimensional universes remains spatiotemporal, unlike in other approaches to quantum gravity, including other interpretations of string theory.\footnote{Brane cosmology, for the technically inclined reader, presupposes an ontology of open strings with endpoints confined to higher-dimensional regions---D-branes---on which gauge interactions are localized. This ontological commitment excludes the heterotic superstring theories from offering an equally literal or insightful account, since they contain only closed strings and lack D-branes \citep{le2023string}. As such, brane cosmology requires a discriminatory stance on duality, challenging the view that all dual descriptions are ontologically equivalent \citep{le2018duality}, and pushing against claims about the emergence of spacetime in string theory that rely on a non-discriminatory approach to duality \citep{HuggettDualities, HUWU}. Accordingly, brane cosmology should not be seen as a case of spacetime emergence in the usual sense, where the emergent base is non-spatiotemporal.} However, quantum gravity theories suggest the possibility that the larger structure within which these universes exist need not itself be spatiotemporal. This theoretical insight thus opens the possibility of non-spatiotemporal causal interactions between universes. This scenario introduces the concept of \emph{non-spatiotemporal causal signature}, broadening our conception of what empirical access might involve. Such access would transcend spacetime while remaining causal and---possibly and to some extent---direct. This perspective suggests that the empirical signature of multiverse scenarios could, at least in principle, arise through causal structures extending beyond the standard spatiotemporal causation.

\section{Generalizing and Addressing the `This Universe' Objection}\label{objection}

The main question of this chapter is whether our universe could provide empirical support for multiverse scenarios. A natural challenge to that idea is what I call the `epistemic isolation objection': the claim that since universes are physically and metaphysically isolated, they are also epistemically isolated. In other words, the nature and existence of other universes, or of a multiversal structure, cannot, as a matter of principle, determine empirically accessible data inside our own universe.

This objection generalizes the ‘this universe’ objection (TUO) of \citet{hacking1987inverse} and \citet{white2003fine}, which in its standard form targets multiverse-based explanations of fine-tuning. The argument is that the existence of other universes with different parameter values is absolutely independent of the values in the particular universe we inhabit, in such a way that positing a multiverse cannot explain why ‘this’ universe appears fine-tuned for life. Thus, as usually formulated, the argument tacitly assumes that the constants of our universe are physically brute facts and independent of any wider explanatory physical structure. Although I have expressed skepticism about the probative weight of fine-tuning arguments (see Section 2), I do not think the TUO successfully undercuts them.

First, I agree with \citet{Ruyant2025-RUYIBB} that the TUO fails. As Ruyant explains, the objection mistakenly treats `this universe' as rigidly referring to a particular universe, overlooking the selection effect involved in our finding ourselves in a life-permitting universe.

Second, another reply is offered by \citet{manson2003fine}. They argue that the TUO presupposes that parameter values are metaphysically contingent, whereas in physical practice they are treated as essential to universes. In other words, universes essentially instantiate those quantitative properties. In response, it has been argued that the essentialist diagnosis is not essential to the objection. For example, \citeauthor{BoyceForthcoming-BOYOEA} (2026) develops a formal probabilistic strategy designed to preserve the TUO without relying on essentialist background commitments. 

However, even these more sophisticated formulations of the TUO inherit two tacit presuppositions: (i) independence of parameter values, i.e, the constants of our universe are probabilistically independent of what happens in other universes (since there is no mechanism connecting them); and (ii) arbitrariness, i.e. the values of those parameters are not determined by a wider structure. Once these presuppositions are denied, as they can be in scientific multiverse practice, where universes are generated or constrained by a common mechanism, the TUO dissolves. This is why the whole debate remains confined to what I call the \textit{`this universe objection' a priori battlefield}: intricate disputes over how Bayesian treatments of our evidence interact with fine-tuning arguments.\footnote{See, e.g., Friederich (2021).} 

What I propose instead is to build on and refine Manson and Thrush’s essentialist insight by making explicit the explanatory mechanism that gives rise to the essentialist belief. The central idea is this: physicists introduce multiverse hypotheses as broader structures that determine the parameter values or features of individual universes. Universes are not isolated islands with brute, arbitrary constants: their specific values are embedded in and explained by a broader framework ontologically committed to determination relations between the multiverse and the universes. The essentiality of those parameters is then only an indirect byproduct of that metaphysical relation of determination (causation or grounding).

Both the TUO and epistemic isolation objections rest on a shared denial: they reject that there is any substantive explanatory or metaphysical determination relation between the multiverse and the universes it contains. But, as I have argued, in scientific practice, deploying multiverse hypotheses almost always presupposes some form of ontic determination from the multiverse to the parameters or features of a given universe. In other words, scientific multiverses are virtually always holistic multiverses.\footnote{As we have seen in the previous section, the string theory landscape, taken in isolation, might be the only counterexample: without coupling to a concrete cosmological mechanism, it can function as a rationally motivated fragmented multiverse.}

Philosophers have often tried to adjudicate the existence of the multiverse a priori, deploying Bayesian arguments. But such Bayesian frameworks tend to conflate idealized a priori coherence with empirical knowledge of the structure of reality. They move too quickly from ‘what rationally ought to be believed’ to ‘what metaphysically is’. In contrast, in the scientific multiverse contexts I am concerned with, multiverse-to-universe determination is part of the physical theory itself.

The contrast between fragmented and holistic multiverses, and the realization that scientific multiverses are almost always holistic, helps explain why, as \citet[40]{manson2020multiverse} notes, scientists never invoke the TUO in working practice. Scientific multiverses are introduced precisely to supply explanatory determination, causal or otherwise, linking the multiverse to individual universes.

\section{Conclusion}\label{sec:conc}

The first two objections often raised against multiverse scenarios, that they posit unobservable entities and that they lack \emph{current} empirical support, were shown to be weak. The more serious challenge is the third: the claim that multiverses could not, even in principle, receive empirical support. This chapter has argued that the force of that objection depends on the \textit{metaphysical profile} of the multiverse in question.

\textit{Fragmented multiverses} lack any intrinsic features that could, even in principle, be empirically accessible from within a single universe. Their architecture is not governed by a unified physical mechanism, but rather posited externally or theologically. Any connection between the universes in such a model is purely non-physical, and can only be defended on a priori grounds. From an empiricist perspective, such fragmented multiverses are effectively invisible: they cannot explain observable features of our world and fall outside the scope of confirmation. However, they might still be scientific, in an unexpected way.

As Christian Wüthrich argues, if spacetime emerges from a non-spatiotemporal structure in quantum gravity, then David Lewis’s modal realism, understood as a kind of fragmented multiverse, conflicts with the actual structure of reality \citep{wuthrich2018actual}. For Lewis, spatiotemporal relations (or their analogues) are world-making relations; if spacetime is not fundamental, these relations cannot straightforwardly serve their intended role. Thus, even though fragmented multiverses cannot leave an empirical signature within our universe, one can still learn that a once-legitimate scenario is inconsistent with the best available physics. In this sense, at least some fragmented multiverse scenarios can be empirically falsified.

\textit{Holistic multiverses}, by contrast, posit a unifying structure---such as a universal quantum state, an inflating field, or a pattern of dynamical laws---that governs the generation or character of the universes they contain. This structure is explanatory in a way that can ground empirical features within individual universes. Scientific multiverse proposals typically assume such grounding relations. They are not merely frameworks based on purely a priori speculation, but ones that offer testable consequences, explanatory power, and integration with our best physical theories. Their potential confirmation or refutation could proceed via direct causal traces (e.g., bubble collisions), local grounding (e.g., interference in EQM), or global grounding (e.g., via statistical patterns in EQM). In addition, a multiverse hypothesis could be indirectly empirically supported through the direct theoretical predictions from empirically supported physical theories, as illustrated by eternal inflation.

Sometimes, holistic multiverses admit empirical confirmation in principle, but are more resistant to falsification. For instance, in eternal inflation, the \emph{measure problem} makes ensemble-level predictions highly flexible, while the absence of bubble-collision traces in the CMB does not count as strong disconfirmation (since such traces are not predicted by all models). However, empirical falsification remains possible in other scenarios. In EQM, for example, the detection of an empirical signature distinctively associated with a spontaneous collapse theory of quantum mechanics could serve as a decisive refutation of unitary quantum mechanics in general, and thus of EQM in particular.

In short, the epistemic profiles are distinct: fragmented multiverses are closed to confirmation through multiversal signatures, but some remain open to falsification through incompatibility with physics, whereas holistic multiverses can admit both confirmation and refutation. The `this universe' and epistemic isolation objections therefore retain their force against fragmented multiverses, but not against holistic multiverses. Once determination relations are acknowledged, those objections collapse, and empirical access becomes possible. The stronger claim that multiverse hypotheses are inherently and universally untestable, and therefore unscientific, is therefore misguided. At least some multiverse scenarios are open to empirical investigation, while others remain subject to indirect constraint via incompatibility with physics. This helps to explain why physicists themselves rarely engage with the `this universe' objection: in practice, multiverse scenarios are introduced as explanatory frameworks embedding universes in broader structures that allow for empirical signatures.

\end{document}